\documentclass[twocolumn,prd,showpacs]{revtex4}
\usepackage{graphicx}
\begin{document}

\title{%
  Dark matter annihilation: the origin of cosmic gamma-ray 
  background at 1--20~MeV
}%

\author{Kyungjin Ahn}
\author{Eiichiro Komatsu}

\affiliation{
Department of Astronomy, University of Texas at Austin\\
1 University Station, C1400, Austin, TX 78712
}

\begin{abstract}
  The origin of the cosmic $\gamma$-ray background at 1--20~MeV 
remains a mystery. We show that $\gamma$-ray emission accompanying 
annihilation of 20~MeV dark matter particles explains most of the 
observed signal. Our model satisfies all of the current observational 
constraints, and naturally provides the origin of ``missing'' $\gamma$-ray 
background at 1--20~MeV and 511~keV line emission from the Galactic center.
We conclude that $\gamma$-ray observations support the existence
of 20~MeV dark matter particles. Improved measurements of the $\gamma$-ray 
background in this energy band undoubtedly test our proposal.
\end{abstract}
\pacs{95.35.+d, 95.85.Nv, 95.85.Pw}

\maketitle

What is the origin of the cosmic $\gamma$-ray background?
It is usually understood that the cosmic $\gamma$-ray background
is a superposition of unresolved astronomical $\gamma$-ray sources 
distributed in the universe.
Active Galactic Nuclei (AGNs) alone explain most of the background light
in two energy regions: 
ordinary (but obscured by intervening hydrogen gas) AGNs account for 
the low-energy ($\lesssim 0.5$~MeV) spectrum
\cite{comastri,zdziarski/etal:1995,ueda}, 
whereas
beamed AGNs (known as Blazars) account for
the high-energy ($\gtrsim 20~{\rm MeV}$) spectrum
\cite{salamon/stecker:1994,stecker/salamon:1996,pavlidou/fields:2002}.
There is, however, a gap between these two regions.
While historically supernovae have been a leading candidate for 
the background up to 4~MeV
\cite{clayton/ward:1975,the,zdziarski:1996,watanabe/etal:1999},
recent studies\cite{strigari/etal:2005,ahn/komatsu/hoeflich:2005} 
show that the supernova contribution is an order of magnitude lower than
observed. The spectrum at 4--20~MeV also remains unexplained
(for a review on this subject, see \cite{stecker/salamon:2001}).
It is not very easy to explain such high-energy background light
by astronomical sources without AGNs or supernovae.

So, what is the origin of the cosmic $\gamma$-ray background
at 0.5--20~MeV? On energetics, a decay or annihilation of particles 
having mass in the range of $0.5~{\rm MeV}\lesssim m_X \lesssim 20~{\rm MeV}$ 
would produce the background light in the desired energy band. 
Since both lower- and higher-energy spectra are already
accounted for by AGNs almost entirely, too lighter or too heavier 
(e.g., neutralinos) particles should be excluded. 
Is there any evidence or reason that such particles should exist?
The most compelling evidence comes from 511~keV line emission
from the central part of our Galaxy, which has been detected and 
mapped by the SPI spectrometer on the INTErnational Gamma-Ray Astrophysics 
Laboratory (INTEGRAL) satellite\cite{spi,spi2}.
This line should be produced by annihilation of electron-positron pairs, 
and one of the possible origins is the dark matter particles annihilating 
into electron-positron pairs\cite{mev_dm}.
This proposal explains the measured injection rate of positrons as well as
morphology of the signal extended over the bulge region.
Intriguingly, popular astronomical sources such as supernovae again seem to fail 
to satisfy the observational constraints\cite{dwarf}.
Motivated by this idea, in the previous paper\cite{ahn/komatsu:2005} 
we have calculated the $\gamma$-ray background of redshifted 511~keV lines 
from extragalactic halos distributed over a large redshift range. 
We have shown that the annihilation signal makes a substantial contribution 
to the low-energy spectrum at $<0.511$~MeV, which constrains $m_X$ to be 
heavier than 20~MeV in order for the sum of the AGN 
and annihilation contributions not to exceed the observed signal.

In this paper, we extend our previous analysis to include continuum
emission accompanying annihilation.
The emerging continuum spectrum should of course depend on the precise
nature of dark matter particles, which is yet to be determined.
Recently, an interesting proposal was made by \cite{beacom/bell/bertone:2004}:
radiative corrections to annihilation, $XX\rightarrow
e^+e^-$, should lead to emission of $\gamma$-rays via the internal 
bremsstrahlung, the emission of extra final-state photons during a 
reaction, $XX\rightarrow e^+e^-\gamma$. 
They have calculated the spectrum of the internal
bremsstrahlung expected for annihilation in the Galactic center, 
compared to the Galactic $\gamma$-ray data, and obtained a constraint on mass as 
$m_X\lesssim 20$~MeV.
A crucial assumption in their analysis is that the cross section of
internal bremsstrahlung is linearly proportional to the annihilation cross
section, and the constant of proportionality is independent of
the nature of annihilation, as is found for related processes
\cite{crittenden/walker/ballam:1961,martyn,berends/bohm,bergstrom/etal:2005}.
More specifically, they assumed that the cross section of 
$XX\rightarrow e^+e^-\gamma$ would be calculated by
that of $e^+e^-\rightarrow \mu^+\mu^-\gamma$ with the muon mass
replaced by the electron mass.
Although the equivalence between these two processes/cross-sections
has not been demonstrated as yet,  
we adopt their procedure into our calculations.

We calculate the background intensity, $I_{\nu}$, as \cite{peacock}
\begin{equation}
 I_{\nu} =
 \frac{c}{4\pi} 
 \int 
 \frac{dz\, P_{\nu}([1+z]\nu, z)}{H(z) (1+z)^{4}},
\label{eq-inu-generic}
\end{equation}
where $\nu$ is an observed frequency, $H(z)$ is the expansion rate at
redshift $z$, and 
$P_{\nu}(\nu, z)$ is the volume emissivity (in units of energy per unit time, 
unit frequency and unit proper volume):
\begin{equation}
  P_{\nu}=\frac12h\nu \langle \sigma v\rangle n^2_X
  \left[\frac{4\alpha}{\pi}\frac{g(\nu)}{\nu}\right],
\end{equation}
where $\alpha\simeq 1/137$ is the fine structure constant, 
$n_X$ is the number density of dark matter particles,
and
$\langle \sigma v\rangle$ is the thermally 
averaged annihilation cross section.
To fully account for WMAP's determination of mass density of 
dark matter\cite{wmap}, $\Omega_{X}h^2=0.113$, by cold relics
from the early universe, one finds
$\left\langle \sigma v\right\rangle
= [3.9, 2.7, 3.2]\,10^{-26}~\mathrm{cm^{3}~s^{-1}}$ for
$m_{X}=[1,\,10,\,100]~\mathrm{MeV}$, respectively
(e.g., see Eq.~[1] in \cite{bes}).
We have assumed that 
$\langle \sigma v\rangle$ is velocity-independent (S-wave annihilation).
One might add a velocity-dependent term (such as P-wave annihilation)
to the cross-section; however, such terms add more degrees of freedom to 
the model, making the model less predictable.
While B\oe hm et al.\cite{mev_dm} argue that the S-wave cross section overpredicts 
the $\gamma$-ray flux from the Galactic center, 
we have shown in the previous paper\cite{ahn/komatsu:2005}
that it is still consistent with the data for
$m_X\gtrsim 20$~MeV and the Galactic density profile of
$\rho\propto r^{-0.4}$ or shallower. 
(We shall discuss an issue regarding the density profile later.)
Finally, a dimensionless spectral function, $g(\nu)$, is defined by
\begin{equation}
 g(\nu) \equiv \frac14\left(\ln\frac{s'}{m_e^2}-1\right)
 \left[1+\left(\frac{s'}{4m_X^2}\right)^2\right],
 \label{eq:gfunc}
\end{equation}
where $s'\equiv 4m_X(m_X-h\nu)$. This function is approximately
constant for $h\nu < m_X$, and then sharply cuts off at $h\nu \sim m_X$.
Thus, one may approximate it as 
\begin{equation}
 g(\nu)\approx  \ln\left(\frac{2m_X}{m_e}\right)\vartheta(m_X-h\nu)
 \label{eq:gapprox}
\end{equation}
for the sake of an order-of-magnitude estimation.
(Note that we have also assumed $m_X\gg m_e$.)

Since the number density is usually
unknown, we use the mass density, $\rho_X\equiv n_X/m_X$, instead. 
After multiplying by $\nu$, one obtains
\begin{eqnarray}
\nu I_{\nu} 
 \nonumber
 &=& \frac{\alpha h\nu\left\langle \sigma v\right\rangle}{2\pi^2 m_{X}^{2}}
 \int_0^\infty \frac{dz~cg[(1+z)\nu]}{H(z)}
 \left\langle\rho_{X}^{2}\right\rangle_{z} \\
 &\simeq&  
\nonumber
3.800~{\rm keV~cm^{-2}~s^{-1}~str^{-1}} \\
\nonumber
& &\times
\left(\frac{\left\langle\sigma v\right\rangle}{10^{-26}~{\rm cm^{3}~s^{-1}}}\right)
\left(\frac{h\nu~1~{\rm MeV}}{m_X^2}\right)\\
& &\times
\int dz\frac{g[(1+z)\nu](1+z)^2(\Omega_X h^2)^2}{\sqrt{\Omega_{m}h^2(1+z)^3+\Omega_\Lambda h^2}}
\frac{C_X(z)}{10^3},
\label{eq-nuinu}
\end{eqnarray}
where $\left\langle\rho_{X}^{2}\right\rangle_{z}$
is the average of $\rho_{X}^{2}$ over proper volume at $z$,
and $C_X(z)\equiv \left\langle\rho_{X}\right\rangle^2_{z}/\left\langle
\rho_{X}^{2}\right\rangle_{z}$ is the dark matter clumping factor.
(We have used $\left\langle\rho_{X}\right\rangle_z=10.54~\Omega_X h^2(1+z)^3~{\rm keV~cm^{-3}}$.)
While equation~(\ref{eq-nuinu}) is exact, one may obtain a better analytical
insight of this equation by using
the approximation to $g(\nu)$ (Eq.~[\ref{eq:gapprox}]),
\begin{eqnarray}
\nu I_{\nu} 
 &\simeq&  
\nonumber
3.800~{\rm keV~cm^{-2}~s^{-1}~str^{-1}} \\
\nonumber
& &\times
\ln\left(\frac{2m_X}{0.511~{\rm MeV}}\right)
\left[\frac{(\Omega_Xh^2)^2}{\sqrt{\Omega_mh^2}}\right]
\left(\frac{\left\langle\sigma v\right\rangle}{10^{-26}~{\rm cm^{3}~s^{-1}}}\right)
\\
& &\times
\sqrt{\frac{1~{\rm MeV}^2}{h\nu~m_X}}\int_{h\nu/m_X}^1 dy~y^{1/2}\frac{C_X[(m_X/h\nu)y]}{10^3},
\label{eq-nuinuapprox}
\end{eqnarray}
where $y\equiv h\nu(1+z)/m_X$. 
Here, we have also assumed that the integral is dominated by 
$1+z\gg (\Omega_\Lambda/\Omega_m) = 2.3$.

We follow the method developed in our previous paper 
\cite{ahn/komatsu:2005} for calculating the clumping factor of dark matter, 
$C_X(z)$.
We have shown that $C_X(z)$ at $z\lesssim 20$ is approximately a power law, 
\begin{equation}
 C_X(z)= C_X(0)(1+z)^{-\beta}, 
\end{equation}
and
$\beta$ depends on adopted dark matter halo profiles.
For example, a cuspy profile such as the Navarro-Frenk-White (NFW) 
profile \cite{nfw}, $\rho_X(r)\propto r^{-1}$, gives 
$C_X(0)\simeq 10^5$ and $\beta\simeq 1.8$,
while a flat profile such as the Truncated Isothermal Sphere
(TIS) \cite{tis}, $\rho_X(r)\propto r^0$, gives 
$C_X(0)\simeq 10^3$ and $\beta\simeq 0$
(see Figure~2 of \cite{ahn/komatsu:2005}).
Using a power-law evolution of $C_X(z)$, one obtains an approximate
shape of the spectrum as
\begin{equation}
\nu I_\nu
\propto
\frac{h\nu\ln (2m_X/m_e)}{(\beta-3/2)m_X^2}
\left[1-\left(\frac{h\nu}{m_X}\right)^{\beta-3/2}\right]\vartheta(m_X-h\nu),
\label{eq:app}
\end{equation}
for $m_X\gg m_e$.
If $\beta<3/2$ (e.g., TIS), 
$\nu I_\nu\propto (h\nu)^{\beta-1/2}(\ln m_X)/m_X^{\beta+1/2}\vartheta(m_X-h\nu)$,
whereas if $\beta>3/2$ (e.g., NFW), 
$\nu I_\nu\propto h\nu [\ln(2m_X/m_e)]/m_X^2\vartheta(m_X-h\nu)$.
Note that the shape of the spectrum becomes insensitive to halo profiles
for the latter case (while the amplitude still depends on profiles).

\begin{figure}
\includegraphics[width=86mm]{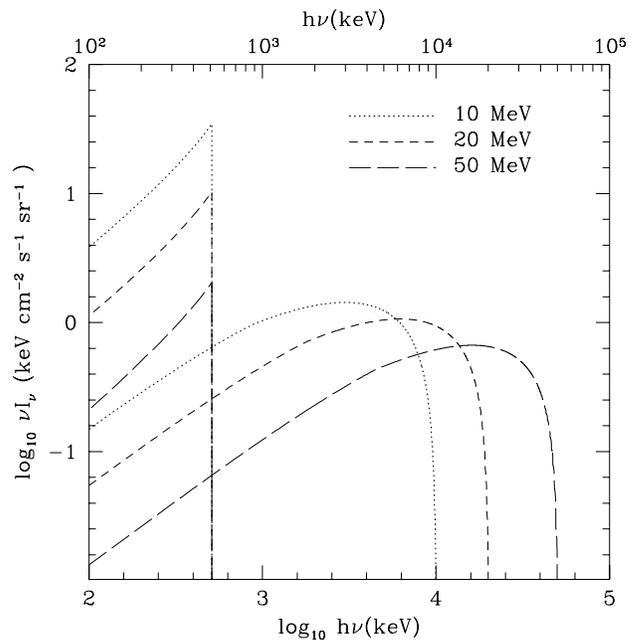}
\caption{\label{fig:spectrum}
  Cosmic $\gamma$-ray background from dark matter annihilation. 
  The dotted, short-dashed, and long-dashed lines 
  show $m_X=10$, 20, and 50~MeV, respectively.
  The curves which sharply cut off at
  511~keV represent background light from line 
  emission\cite{ahn/komatsu:2005}, 
  while the others which extend to higher energy represent the
  internal bremsstrahlung. 
}
\end{figure}

\begin{figure}
\includegraphics[width=86mm]{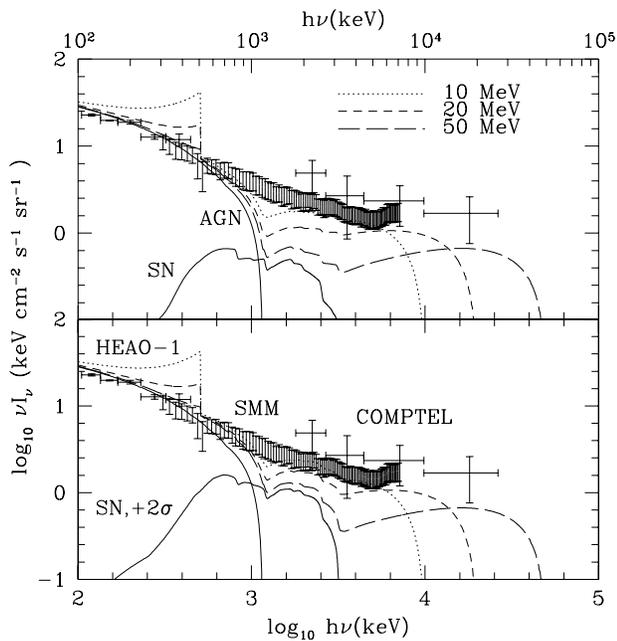}
\caption{\label{fig:spectrum-tot}
  The total cosmic $\gamma$-ray background produced by dark matter
  annihilation, AGNs\cite{ueda}, and Type Ia 
  supernovae\cite{ahn/komatsu/hoeflich:2005}.   
  The dotted, short-dashed, and long-dashed lines 
  show $m_X=10$, 20, and 50~MeV, respectively.
  The supernova contribution depends on the observed supernova rate, 
  and we consider the best-fit rate (upper panel) as well as
  the $2\sigma$ upper limit (lower panel).
  The data points of HEAO-1 A4 MED\cite{heao}, SMM\cite{smm}, and 
  COMPTEL\cite{comptel} experiments are also shown.
}
\end{figure}

\begin{figure}
\includegraphics[width=86mm]{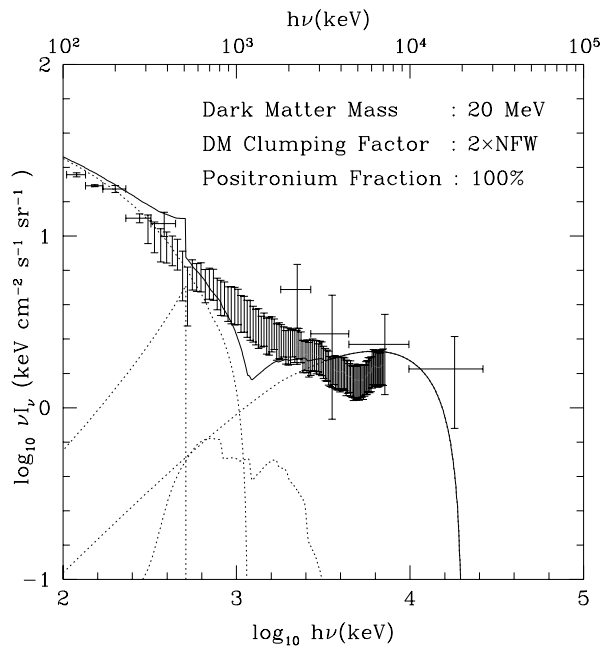}
\caption{\label{fig:spectrum-tot2}
  The best-fit model of the cosmic $\gamma$-ray background. 
  The model assumes
(a) $m_X=20$~MeV, (b) the mean dark matter clumping factor
is twice as large as predicted by the NFW profile
(due to either a steeper profile or the presence of substructures),
and (c) line emission is solely produced via positronium formation.
The dashed lines show each contribution separately.
}
\end{figure}

Henceforth we shall adopt the NFW profile as the fiducial model, as
it fits the mean central halo profiles in numerical simulations well.
Following the previous paper, we take into account a scatter in halo profiles
by integrating over a probability distribution of halo concentration;
thus, our model effectively incorporates significantly less concentrated 
(such as our Galaxy) or more concentrated profiles than the average NFW.  
One might argue that our model based on the NFW profile
is unable to explain $\gamma$-ray emission from the Galactic center,
which requires $\rho\propto r^{-0.4}$ (or shallower).
If desired, one might use this profile and recalculate the $\gamma$-ray
background spectrum; however, we continue to use the NFW profile,
assuming that our Galaxy is not a ``typical'' halo in the universe.
If there are so many more galaxies which obey the NFW profile, then
the signal should be dominated by those typical halos.
Of course, real universe does not have to be the same as 
numerical simulations, and one way to incorporate 
the uncertainty of halo profiles into our analysis would 
be to treat $C_X(0)$ and $\beta$ as free parameters. We shall
come back to this point at the end of this paper.
Figure~\ref{fig:spectrum} shows the predicted cosmic $\gamma$-ray
background from dark matter annihilation, including
line\cite{ahn/komatsu:2005} and continuum emission,
for $m_X=10$, 20, and 50~MeV.
The shape of the internal bremsstrahlung 
is described well by the approximate formula (Eq.~[\ref{eq:app}])
with $\beta=1.8$. As expected, the continuum spectrum extends 
up to $h\nu\sim m_X$, whereas line emission contributes
only at $<0.511$~MeV.

Now let us add extra contributions from known astronomical sources
and compare the total predicted spectrum with the observational data.
Figure~\ref{fig:spectrum-tot} compares the sum of dark matter 
annihilation, AGNs\cite{ueda} 
and Type Ia supernovae\cite{ahn/komatsu/hoeflich:2005}
with the data points of HEAO-1\cite{heao}, SMM\cite{smm}, and COMPTEL\cite{comptel}
experiments.
We find that $m_X\sim 20$~MeV fits
the low-energy spectrum\cite{ahn/komatsu:2005} and
explains about a half of the spectrum at 1--20~MeV.
Therefore, the internal bremsstrahlung from dark matter
annihilation is a very attractive source
of the cosmic $\gamma$-ray background in this energy region.
It is remarkable that such a simple model provides adequate explanations 
to two completely different problems:
511~keV line emission from the Galactic center\cite{mev_dm}, and
missing $\gamma$-ray light at 1--20~MeV.
(The regular Blazars would dominate the spectrum beyond 
20~MeV\cite{salamon/stecker:1994,stecker/salamon:1996,pavlidou/fields:2002}.)

If desired, one might try to improve agreement with the data 
in the following way. 
The continuum (combined with the other contributions)
can fully account for the SMM and COMPTEL data, 
if the clumping factor is twice as large as predicted
by the NFW profile.
This could be easily done within uncertainty in 
our understanding of the structure of dark matter halos: for example, 
a slightly steeper profile, or the presence of substructure\cite{substruct}.
However, a larger clumping factor also increases 511~keV line emission 
by the same amount, 
which would exceed the HEAO-1 and SMM data. How do we reduce line emission 
independent of continuum?  
The line emission is suppressed by up to a factor of 4, 
if $e^+e^-$ annihilation occurs
predominantly via positronium formation.
Once formed, a positronium decays into either two 511~keV photons
or three continuum photons. 
As the branching ratio of the former process is only 1/4,
line emission is suppressed by a factor of 4 if all of annihilation
occurs via positronium formation.
If a fraction, $f$, of annihilation occurs via
positronium, then line is suppressed by 
$1-3f/4$\cite{beacom/bell/bertone:2004};
thus, we can cancel the effect of doubling the clumping by requiring that
2/3 of line emission be produced via positronium.
Figure~\ref{fig:spectrum-tot2} shows our ``best-fit'' model, which
assumes (a) $m_X=20$~MeV, (b) the mean clumping factor is twice as large, 
and (c) line emission is solely produced via positronium ($f=1$).
Note that this is a reasonable extension of the minimal model
and makes the model more realistic: we know from simulations that 
there must exist substructures in halos. Some 
fraction of line emission must be produced via positronium,
as it has been known that more than 90\% of 511~keV emission from the 
Galactic center is actually produced via positronium formation
\cite{kinzer/etal:2001,churazov/etal:2005}.
While the model seems to slightly exceed the HEAO-1 and SMM data at low energy,
we do not take it seriously as the discrepancy would be smaller 
than the uncertainty of the AGN model.
The AGN model presented here assumes 
a high-energy cut-off energy of $E_{\rm cut}=0.5$~MeV\cite{ueda}.
Since current data of AGNs in such a high energy band are
fairly limited, uncertainty in $E_{\rm cut}$ is more 
than a factor of 2. Even a slight reduction in $E_{\rm cut}$ 
would make our model fit the low-energy spectrum.

The best-fit model is consistent with and supported by 
all of the current observational constraints: it fits the 
Galactic $\gamma$-ray emission as well as the cosmic 
$\gamma$-ray emission. It might also account for a small 
difference between theory and the experimental data of the muon 
and electron anomalous magnetic moment\cite{boehm/ascasibar:2004}.
We stress here that, to the best of our knowledge, 
all of these data would remain unexplained otherwise.
There is, however, one potential conflict with a new analysis of 
the SPI data by \cite{ascasibar/etal:2005}, which
shows that a NFW density profile 
does provide a good fit to 511~keV line emission from the Galactic center,
as opposed to the previous analysis by \cite{mev_dm}, which
indicated a shallower profile than NFW. 
This new model would have much higher dark-matter clumping 
and require a substantially (more than an order of magnitude)
smaller annihilation cross-section than
$\left\langle \sigma v\right\rangle
\sim 3\times 10^{-26}~\mathrm{cm^{3}~s^{-1}}$ to fit the Galactic data.
Is our Galaxy consistent with NFW?
This is a rather complicated issue which is still far from settled (e.g., 
\cite{binney/evans:2001,klypin/zhao/somerville:2002}), and
more studies are required to understand the precise shape of density
profile of our Galaxy. If our Galaxy is described by a steep profile
such as NFW, then the dark matter annihilation probably makes a 
negligible contribution to the $\gamma$-ray background, unless dark matter clumping
is significantly increased by substructure\cite{substruct},
compensating a small cross section. On the other hand, if it were
confirmed that our Galaxy has a shallow density profile and the
contribution of the dark matter annihilation to the $\gamma$-ray
background is negligible, it would be difficult to explain the
Galactic $\gamma$-ray signal solely by annihilation of light dark
matter particles.

As shown in Figure~\ref{fig:spectrum-tot2}, dark matter annihilation
produces a distinctive $\gamma$-ray spectrum at 0.1--20~MeV.
More precise determinations of the cosmic $\gamma$-ray background
in this energy band will undoubtedly test our proposal. 
If confirmed, such measurements would shed light on the nature of 
dark matter, and potentially open a window to new physics:
one implication is that neutralinos would be excluded from 
a candidate list of dark matter.
Phenomenologically, our model may be parameterized by four free 
parameters: (1) dark matter mass, $m_X$, (2) a dark matter clumping
factor at present, $C_X(0)$, (3) redshift evolution of clumping, $\beta$, and 
(4) a positronium fraction, $f$. 
When more precise data are available in the future, it might be 
possible to perform a full likelihood analysis and constrain
properties of dark matter particles as well as dark matter halos.

Finally, the angular power spectrum of anisotropy of the $\gamma$-ray background at
1--20~MeV would also offer a powerful diagnosis of the detected signal
(see \cite{zhang/beacom:2004} for the contribution from Type Ia supernovae).  
Our model predicts that the angular power spectrum should be given by the trispectrum
(the Fourier transform of the four-point correlation function)
of dark matter halos projected on the sky, as the signal is proportional to 
$\rho^2$. More specifically, the power spectrum should follow precisely that
of the dark matter clumping factor. More high-quality data of the 
cosmic $\gamma$-ray background in this energy band are seriously awaited. 

We would like to thank D.E. Gruber for providing us with the HEAO-1 
and COMPTEL data, K. Watanabe for providing us with the SMM data,
Y. Ueda for providing us with the AGN predictions,
C. B\oe hm for sharing her results on the modeling of SPI data with us,
and C. B\oe hm and J. Beacom for valuable comments on early versions of 
this paper.
We would also like to thank G. Bertone and P.R. Shapiro for discussion.
K. A. was partially supported by NASA Astrophysical Theory Program 
grants NAG5-10825, NAG5-10826, NNG04G177G, and Texas Advanced Research 
Program grant 3658-0624-1999.

\emph{Note added in proof} -- A recent article by Rasera et
al. (\cite{rasera/etal:2005}) argues that our predictions for the
$\gamma$-ray background from the redshifted 511 keV line
(\cite{ahn/komatsu:2005}) were too large because annihilation of
electrons and positrons cannot take place in halos less massive than
$\sim 10^{7} M_{\odot}$, in which baryons cannot collapse. While this
effect reduces the intensity of the line contribution, it does not
affect the continuum emission (i.e., the internal bremsstrahlung), as
the continuum emission is produced before annihilation. Since the
major contribution to the $\gamma$-ray background at 1-20 MeV comes
from the continuum emission, our conclusion in this paper is not
affected by the results in \cite{rasera/etal:2005}.


\end{document}